\documentclass{emulateapj}
\usepackage{natbib,amsmath}

\newcommand{\cmjj}{\mbox{${\rm cm^{-2}}$}}
\newcommand{\hI}{\mbox{${\rm H\,I}$}}
\newcommand{\nhi}{\mbox{$N({\rm H\,I})$}}
\newcommand{\lya}{\mbox{${\rm Ly}\alpha$}}

\newcommand{\apg}{\gtrsim}
\newcommand{\apll}{\lesssim}
\newcommand{\fesc}{\mbox{$f_{\rm esc}$}}
\newcommand{\tll}{\mbox{$\tau_{\rm LL}$}}
\newcommand{\etal}{\ensuremath{\mbox{et~al.}}}

\shorttitle{Escape Fraction in GRB Host Galaxies}
\shortauthors{Chen \etal}

\begin{document}

\slugcomment{Submitted to the Astrophysical Journal Letters}

\title{A New Constraint on the Escape Fraction in Distant Galaxies Using $\gamma$-ray Burst Afterglow Spectroscopy}
\author{Hsiao-Wen Chen\altaffilmark{1}, Jason X.\ Prochaska\altaffilmark{2,3}, and Nickolay Y.\ Gnedin\altaffilmark{4,1,5}}
\altaffiltext{1}{Department of Astronomy \& Astrophysics, University of Chicago, Chicago, IL 60637, USA, {\tt hchen@oddjob.uchicago.edu}}
\altaffiltext{2}{University of California Observatories - Lick Observatory, University of California, Santa Cruz, CA 95064, USA; {\tt xavier@ucolick.edu}} 
\altaffiltext{3}{Department of Astronomy and Astrophyics, University of California, Santa Cruz, CA 95064, USA}
\altaffiltext{4}{Particle Astrophysics Center, Fermi National Accelerator Laboratory, Batavia, IL 60510, USA; {\tt gnedin@fnal.gov}}
\altaffiltext{5}{Kavli Institute for Cosmological Physics, University of Chicago, Chicago, IL 60637, USA}
\begin{abstract}

We describe a new method to measure the escape fraction \fesc\ of
ionizing radiation from distant star-forming galaxies using the
afterglow spectra of long-duration $\gamma$-ray bursts (GRBs).
Optical spectra of GRB afterglows allow us to evaluate the optical
depth of the host ISM, according to the neutral hydrogen column density 
\nhi\ observed along the sightlines toward the star-forming regions where 
the GRBs are found.  Different from previous effort in searching for 
faint, transmitted Lyman continuum photons, our method is not subject to 
background subtraction uncertainties and does not require prior knowledge 
of either the spectral shape of the host galaxy population or the IGM 
\lya\ forest absorption along these GRB sightlines.  Because most GRBs 
occur in sub-$L_*$ galaxies, our study also offers the first constraint 
on \fesc\ for distant low-mass galaxies that dominate the cosmic 
luminosity density.  We have compiled a sample of 28 GRBs at redshift 
$z\apg 2$ for which the underlying \nhi\ in the host ISM are known.  
These GRBs together offer a statistical sampling of the integrated optical 
depth to ionizing photons along random sightlines from star-forming 
regions in the host galaxies, and allow us to estimate the mean escape 
fraction $\langle\fesc\rangle$ averaged over different viewing angles.  
We find $\langle\fesc\rangle=0.02\pm 0.02$ and place a 95\% c.l.  upper 
limit $\langle\fesc\rangle \le 0.075$ for these hosts.  We discuss 
possible biases of our approach and implications of the result.  
Finally, we propose to extend this technique for measuring 
$\langle\fesc\rangle$ at $z\sim 0.2$ using spectra of core-collapse 
supernovae.

\end{abstract} 


\section{Introduction}

Observations of distant QSOs indicate that the intergalactic medium
(IGM) became fully ionized by redshift $z\sim 6$ (e.g.\ Fan, Carilli,
\& Keating 2006).  While at $z<3$ QSOs are the dominant sources of the
ultraviolet background radiation (e.g.\ Haardt \& Madau 1996), at
higher redshifts where the number density of QSOs declines steeply
toward earlier epochs (e.g.\ Willott \etal\ 2005; 
2005; Richards \etal\ 2006) additional ionizing sources are necessary.
The spectral shape of the ultraviolet background radiation inferred
from intervening metal absorption line studies (e.g.\ Haehnelt \etal\
2001) and intergalactic He\,II absorption spectra (e.g.\ Shull \etal\
2004; Reimers \etal\ 2005) suggest that young stars may provide the
dominant ionizing sources during early epochs.

The escape fraction of ionizing radiation, \fesc, specifies the
fraction of stellar-origin ionizing photons ($h\nu > 1$ Ryd) that
escape the interstellar medium (ISM) of star-forming galaxies.
Accurate measurements of \fesc\ are important for quantifying the
relative contribution of ionizing photons to the ultraviolet
background radiation between galaxies and AGN.  For the Milky Way,
estimates based on diffuse H$\alpha$ emission of High Velocity Clouds
(HVC) yield an upper limit of $\fesc \apll 6$\% (e.g.\ Bland-Hawthorn
\& Maloney 1999 and see Weiner \etal\ 2001 for a review).  In the
nearby universe, early observations of starburst galaxies place
constraints at $\fesc\apll 6$\% (e.g.\ Heckman \etal\ 2001).  Bergvall
\etal\ (2006) have reported the first positive detection of Lyman
continuum photons in the spectra of the starburst galaxy Haro 11 using
the {\it Far-UV Space Explorer}.  Their analysis suggests an escape
fraction in this galaxy of $\fesc=1-10$\%.  This finding has, however,
been challenged by Grimes \etal\ (2007), who cannot confirm the
detection of Lyman continuum in the same data set.

At higher redshift, a range of values are reported from $\fesc<6$\% at
$z=1.1-1.4$ (Malkan \etal\ 2003; Siana \etal\ 2007), to between mean
values of no more than 8\% (Giallongo \etal\ 2002; Fern\'andez-Soto
\etal\ 2003; Shapley \etal\ 2006) and $\fesc\approx 13-38$\% (Inoue
\etal\ 2005) at $z\sim 3$, to $\fesc\approx 22$\% at $\langle
z\rangle=3.4$ (Steidel \etal\ 2001)\footnote{Note that many of the
previous publications reported measurements for $f_{\rm esc,rel}$,
which is defined as the ratio of escaped ionizing photons at 912 \AA\
to the observed flux density at 1500 \AA.  Here we have converted
these relative measurements to \fesc\ at 1 Ryd based on their
respective $\langle E(B-V)\rangle$ reported by the authors.}.  In
addition, direct detections of Lyman continuum photons are reported
for two $z\sim 3$ galaxies by Shapley \etal\ (2006), implying
$\fesc\approx 13-20$\% for these two sources.  The large scatter in
the reported $\fesc$ may imply a large variation in the optical depth
across different lines of sight toward the inner regions of the
galaxies (e.g.\ Gnedin, Kravtsov, \& Chen 2007), but it also
underscores the challenges in detecting low-luminosity features in a
background noise limited regime.

In this {\it Letter}, we describe a novel approach for constraining
\fesc\ from high-redshift star-forming galaxies.  We estimate \fesc\
based on the distribution of neutral hydrogen column density \nhi\
observed in the afterglow spectra of long-duration $\gamma$-ray bursts
(GRBs).  Long-duration GRBs are believed to originate in the death of
massive stars with $M > 20$ M$_\odot$ (see Woosley \& Bloom 2006 for a
recent review), and are signposts of active star-forming regions in
the ISM of their host galaxies (e.g.\ Bloom \etal\ 2002; Fruchter
\etal\ 2006).  Spectroscopic observations of the bright optical
afterglows following the initial bursts have allowed us to measure the
gas and dust content along the sightlines toward the GRBs, based on
absorption features imprinted in the afterglow spectra (e.g.\
Vreeswijk \etal\ 2004; Chen \etal\ 2005; Prochaska \etal\ 2007a).
These GRB sightlines together offer a statistical sampling of the
integrated optical depth to ionizing photons along random directions
in the host galaxies.

The escape fraction \fesc\ determined from the \nhi\ distribution
along random sightlines in the ISM of GRB host galaxies does not
require direct detection of Lyman continuum photons.  It is not
subject to systematic uncertainties due to background subtraction.
The \fesc\ value is derived based on the total gas column observed in
front of the star-forming region that hosts the GRB.  It does not
depend on the spectral shape of the ultraviolet radiation from the
host galaxy or the stochastic uncertainties in the IGM \lya\ forest
absorption along the lines of sight.  Finally, while some GRB host
galaxies are reported to have high star formation rate, SFR $\apg 100$
M$_\odot$, (c.f.\ Berger \etal\ 2003; Le Floc'h \etal\ 2006), growing
evidence indicates that the majority are sub-$L_*$ galaxies (e.g.\ Le
Floc'h \etal\ 2003; Sollerman \etal\ 2005; Fruchter \etal\ 2006).  Our
study therefore offers the first constraint on \fesc\ for low-mass
galaxies at $z>2$.


\section{The Sample of GRBs at $z\apg 2$}

To obtain an accurate estimate of the mean escape fraction of ionizing
photons along GRB sightlines, we first compile a sample of GRBs that
are confirmed at $z\apg 2$.  We focus our analysis on bursts at
$z_{\rm GRB} \apg 2$ for two main reasons.  First, it minimizes the
ambiguity between an optically thin sightline and a low-redshift
interloper.  The $z>2$ IGM will imprint its signature on an optical
spectrum via the \lya\ forest and metal-line absorption features.
Therefore, an afterlow that exhibits a featurless spectrum is most
likely at $z_{\rm GRB} < 2$.  Second, measurements of \nhi\ from the
absorption profiles of \lya\ and Lyman series allow us to directly
evaluate the optical depth at the Lyman limit frequency \tll.

Table 1 lists 40 spectroscopically confirmed GRBs at $z_{\rm GRB}\apg
2$.  In addition to the redshift of each source, we also list the
isotropic equivalent energy release in $\gamma$-ray photons ($E_{\rm
iso}$), the observed \nhi\ if available, a flag f$_{\alpha}$ to
indicate whether the afterglow spectrum covers the redshifted \lya\
transition ('0' means no coverage and '1' means \lya\ coverage), and a
flag f$_{i}$ to indicate whether metal-line features due to low ions
such Si\,II or C\,II are present.  Four of the GRBs do not have
spectral coverage of the \lya\ transition from the host, but the
presence of low ions (f$_{i}$) indicate that the gas is consistent
with being optically thick.  To exhibit strong low-ion absorption, an
optically thin gas would need to have very high density (to maintain a
non-negligible neutral fraction) and super-solar metallicity, both of
which are very unlikely.  Nine sources do not have published \nhi.  In
the subsequent analysis, we consider only those 28 sightlines with
published \nhi\ values as our main sample, and assume that the
remaining 13 sources with no available \nhi\ measurements share the
same distribution as the main sample.  This is justified based on the
similar $z_{\rm GRB}$ and $E_{\rm iso}$ distributions between GRB
sightlines with and without known \nhi\ measurements.

Figure 1 presents the cumulative \nhi\ distribution, ${\cal F}[<\nhi]$
from the main sample of 28 GRB host galaxies, together with the
1-$\sigma$ uncertainties determined based on a bootstrap re-sampling
method.  Specifically, we establish a simulated sample of 28
sightlines from random sampling of the main sample, allowing
duplications of individual sightlines.  Then, we evaluate the
cumulative \nhi\ distribution of the simulated sample.  We repeat the
procedure 10,000 times to determine the 68\% scatter of ${\cal F}$
around the mean value in each \nhi\ bin.

\begin{figure}
\begin{center}
\includegraphics[scale=0.35, angle=270]{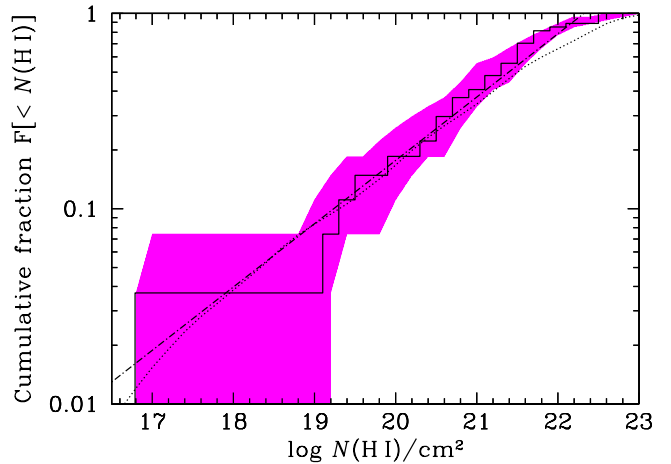}
\caption{Cumulative distribution of neutral hydrogen column density
${\cal F}[<\nhi]$ observed in the host galaxies of long-duration GRBs
at $z\ge 2$ (solid histogram).  The shaded area shows the 1-$\sigma$
uncertainties evaluated using a bootstrap re-sampling method that
accounts for both \nhi\ measurement uncertainties and sampling errors.
The dash-dotted line represents the best-fit power-law model described
in \S\ 3.  The dotted curve respresents the predicted distribution
from Gnedin, Kravtsov, \& Chen (2007).}
\end{center}
\end{figure}

\section{The Method}

The optical depth of Lyman limit photons along individual lines of
sight is determined according to $\tll=\sigma_{\rm LL}\times \nhi$,
where $\sigma_{\rm LL}=6.28\times 10^{-18}$ cm$^2$ is the
photo-ionization cross section of hydrogen atoms.  In principle, {\it
considering a sample of random sightlines from the star-forming
regions in a galaxy together yields an estimate of the optical depth
averaged over all viewing angles}.  In practice, we consider an
ensemble of random sightlines toward GRBs in distant star-forming
galaxies.  The mean escape fraction of Lyman limit photons 
averaged over all directions is evaluated according to
\begin{equation}
\langle\fesc\rangle=\frac{1}{n}\sum_{i=1}^{i=n}\exp[-\sigma_{\rm LL}\,N_i({\rm H\,I})],
\end{equation}
where the sum extends over the total number of $n$ GRB sightlines in
the sample.  For our main sample presented in Figure 1, we find $n=28$
and Equation (1) yields $\langle\fesc\rangle=0.02\pm 0.02$.  The error
is estimated using the bootstrap re-sampling method described in \S\ 2
and represents the 68\% uncertainty in the mean value.  We also
determine a 95\% c.l. upper limit $\langle\fesc\rangle \le 0.075$.

Parameterizing the cumulative \nhi\ distribution by ${\cal F}[<\nhi] =
A\,[\nhi/N_0]^\alpha$, we find based on a $\chi^2$ analysis $\log\,A =
-0.58 \pm 0.05$ and $\alpha = 0.32\pm 0.03$ for $\log\,N_0 = 20.5$
over $\log\,\nhi=16.5-21.5$ (dash-dotted line in Figure 1).  Equation
(1) is expressed as
\begin{eqnarray*}
\langle\fesc\rangle&=&\int_0^\infty d\,N_{\rm H\,I}\frac{d\,{\cal
F}}{d\,N_{\rm H\,I}}\exp[-\sigma_{\rm LL}\,N({\rm H\,I})] \\
&=&\frac{A\,\alpha}{(\sigma_{\rm LL} N_0)^\alpha} \Gamma(\alpha).
\end{eqnarray*}
We determine $\langle\fesc\rangle = 0.020\pm 0.003$, where the errors
represent the 68\% uncertainties.  

\section{Discussion}

We have applied the \nhi\ measured from the \lya\ (and in some cases
Lyman series as well) absorption strength in early-time afterglow
spectra to constrain the mean escape fraction in distant star-forming
galaxies.  Different from conventional methods to search for
transmitted Lyman continuum photons, our estimated
$\langle\fesc\rangle$ is not subject to background subtraction
uncertainties and does not depend on the intrinsic ultraviolet
spectral shape of the host galaxies or dust distribution in the host
ISM.  In addition, it does not depend on the stochastic IGM \lya\
absorption along the sightlines toward these GRBs.  
Here we discuss possible biases in our estimated $\langle\fesc\rangle$
due to the selection of GRB sightlines and implications of our result.

\subsection{Observational Biases}

Our analysis considers only galaxies that host a GRB event.  The
presence of a GRB indicates that the ISM immediately surrounding the
burst is photo-ionized by the bright afterglow.  The intensity of the
afterglow ionizing radiation is capable of photo-ionizing all \hI\ gas
to a distance of $r = 10$ to 30\,pc (Drain \& Hao 2002; Watson \etal\
2007; Prochaska \etal\ 2007b) that can exceed the typical Stromgren
radius of an H\,II region.  It is conceivable that the presence of a
GRB may reduce \tll\ along the sightline and the constraint on
$\langle\fesc\rangle$ can be considered as an upper limit.

In addition, our constraint is derived explicitly at the 912-\AA\
Lyman limit transition.
atoms.  Additional absorption due to dust would further reduce the
estimated $\langle\fesc\rangle$, but is almost negligible at
frequencies beyond 1 Ryd (see Gnedin \etal\ 2007).  However, some
fraction of the GRB afterglows are optically ``dark'' and missed in
our sample because afterglow spectroscopy was impossible.  While the
majority of these are presumably associated with high-$z$ events
(where the IGM absorbs most of the optical photons) or intrinsically
faint afterglows, the remainder are associated with highly
dust-extinguished sightlines (e.g.\ Pellizza \etal\ 2006; Rol \etal\
2007).  These extremely dusty sightlines are optically thick to
ionizing radiation further strengthening the $\langle\fesc\rangle$.

Intrinsically faint UV afterglows could occur in a particularly
low-density environment (Kumar \& Panaitescu 2000) and, in turn,
preferentially probe optically thin sightlines.  We note, however,
that there is no notable correlation between the observed \nhi\ and
the afterglow UV luminosity.  For example, GRB\,021004 and GRB\,050820
have very different \nhi\ but comparable afterglow UV luminosity
(e.g.\ Lazzati \etal\ 2006; Vestrand \etal\ 2006).  Furthermore, the
isotropic equivalent energies of the GRB sample ($E_{\rm iso}$ in
Table 1) display a wide dispersion at all redshift, supporting that
these GRBs do not preferentially originate in low-density environment
(e.g.\ Piran 1999).  We argue that the pre-selection of bright UV
afterglows has not restricted the analysis to a special sub-sample of
GRBs.

\subsection{Implications and Future Work}

The association between GRBs and massive stars that typically have a
short lifetime indicates that the GRB events occur close to the
locations where their progenitor stars were formed and directly trace
the current star formation rate.  Late-time imaging surveys to search
for the host galaxies of long-duration GRBs have shown that these GRBs
occur in sub-$L_*$, galaxies (Le Floc'h \etal\ 2003; Sollerman \etal\
2005; Fruchter \etal\ 2006) that exhibit on average large specific
star-formation rates, SFR/$L_B$, (Christensen \etal\ 2004).  This can
be understood as low-mass galaxies undergoing early generations of
star formation (Erb \etal\ 2006).  The derivation of \fesc\ from GRB
sightlines reveals the escape fraction during the lifetime of the most
intense ionizing sources, and applies to `normal', star-forming
galaxies that dominates the cosmic UV luminosity density.  The
low-mass nature also allows a direct comparison with predictions from
high-resolution cosmological simulations (the dotted curve in Figure
1; Gnedin \etal\ 2007).

Our 95\% c.l. upper limit of $\langle\fesc\rangle$ is comparable to
the low values previously reported based on observations of Lyman
continuum photons from more luminous star-forming galaxies at $z\sim
3$ (Giallongo \etal\ 2002; Fern\'andez-Soto \etal\ 2003; Shapley
\etal\ 2006).  Additional absorption due to dust would further reduce
the estimated $\langle\fesc\rangle$, but is almost negligible at
frequencies beyond 1 Ryd (see Gnedin \etal\ 2007).  We estimate the
total contribution of ionizing photons from sub-$L*$ galaxies,
adopting the UV luminosity function determined for luminous
star-forming galaxies at $z\sim 3$ from Adelberger \& Steidel (2000).
We derive a co-moving luminosity density at 1500 \AA\ of $2.2\times
10^{26}\, h$ erg s$^{-1}$ Hz$^{-1}$ Mpc$^{-3}$ for sub-$L_*$
($0.1-1\,L_*$) galaxies\footnote{We adopt a $\Lambda$ cosmology,
$\Omega_{\rm M}=0.3$ and $\Omega_\Lambda = 0.7$, with a dimensionless
Hubble constant $h = H_0/(100 \ {\rm km} \ {\rm s}^{-1}\ {\rm
Mpc}^{-1})$.}.  Applying an extinction correction to the observed
1500-\AA\ flux (the authors estimated $f_{\rm esc}(1500 \AA)\approx
0.2$) and assuming an intrinsic flux ratio between rest-frame 1500
\AA\ and 900 \AA\ of $f(1500)/f(900)=3$ adopted by Steidel \etal\ 2001
(but see Siana \etal\ 2007), we estimate a co-moving emissivity at 1
Ryd from sub-$L_*$ galaxies of $<2.8\times 10^{25}\, h$ erg s$^{-1}$
Hz$^{-1}$ Mpc$^{-3}$ for the 95\% c.l. upper limit
$\langle\fesc\rangle \apll 0.075$.  To estimate the QSO contribution
to the ultraviolet background radiation, we adopt 
the $z=3$ QSO luminosity function estimated by Hopkins \etal\ (2007).
We find that the contribution to the ionizing background from QSOs of
bolometric luminosity $L_{\rm bol} > 10^8\,L_\odot$ is $\approx
5\times 10^{24}\, h$ erg s$^{-1}$ Hz$^{-1}$ Mpc$^{-3}$.  While the
uncertainties in these various numbers are large, this exercise shows
that QSOs and sub-$L*$ galaxies with $\langle\fesc\rangle=1-2$ \% can
contribute a comparable amount of ionizing photons to the ultraviolet
background radiation at $z\sim 3$.

A larger sample of GRB sightlines with known \nhi\ measurements for
the host ISM is needed for improving the uncertainties in
$\langle\fesc\rangle$.  In addition, follow-up imaging surveys for
unvailing the emission properties of the GRB host galaxies are
valuable for testing model predictions of star formation at high
redshift.  Specifically, cosmological simulations show that low-mass
galaxies are inefficient in emitting ionizing radiation (Gnedin \etal\
2007).  The luminosity distribution of these GRB hosts offers a direct
test of these models.

Finally, it is worth exploring whether a similar analysis could be
performed with core-collapse supernovae (SN).  These events also trace
the death of massive stars over a broader mass range and provide a
bright probe of the optical depth through the host galaxy.
While the extreme line-blanketing in the far-ultraviolet of SN spectra
may preclude the study of Hydrogen absorption, it may be plausible to
pursue an analysis of ISM metal-absorption, e.g.\ via the Mg\,II
doublet.  Indeed, the {\it HST} spectrum of the Type\,II SN 1999em
shows narrow absorption lines of Fe\,II, Mg\,II, and Mg\,I (Baron
\etal\ 2000).  These features imply that NGC\,1637 is optically thick
to ionizing radiation along this particular sightline.  A survey of $z
\gtrsim 0.2$ SN could be carried out with a modest resolution,
blue-sensitive spectrometer on a 10~m-class telescope.

\acknowledgements

The authors acknowledge helpful discussion with J.\ O'Meara, E.\
Ramirez-Ruiz and M.\ Dessauges-Zavadsky.  We thank A.\ Kann and J.\
Bland-Hawthorn for valuable input.  H.-W.C. acknowledges support from
NASA grant NNG\,06GC36G and an NSF grant AST-0607510.
J. X. P. acknowledges support from NASA/Swift grant NNG\,05GF55G and a
CAREER grant (AST-0548180).

\clearpage

\begin{deluxetable}{lccrccr}
\tablecaption{The Sample of GRBs at $z\ge 2$}
\tablewidth{0pt}
\tablehead{\colhead{GRB} & \colhead{$z_{\rm GRB}$} & 
\colhead{$\log\,E_{\rm iso}$\tablenotemark{a}} & \colhead{$\log\,N(\hI)$} & 
\colhead{f$_\alpha$\tablenotemark{b}} & \colhead{f$_i$\tablenotemark{c}} & 
\colhead{Ref.}}
\startdata
000301c & 2.03 & 52.64 & $21.2\pm 0.5$   & 1 & 7 & 1 \nl
000926  & 2.04 & 53.76 & $21.30\pm 0.25$ & 1 & 7 & 2 \nl
011211  & 2.14 & 53.17 & $20.4\pm 0.2$   & 1 & 7 & 3 \nl
020124  & 3.20 & 52.81 & $21.7\pm 0.4$   & 1 & 7 & 33 \nl 
021004  & 2.33 & 52.82 & $19.5\pm 0.5$   & 1 & 3 & 4 \nl
030226  & 1.99 & 53.20 & $20.5\pm 0.3$   & 1 & 7 & 34 \nl
030323  & 3.37 & 52.99 & $21.90\pm 0.07$ & 1 & 7 & 5 \nl
030429  & 2.65 & 52.87 & $21.6\pm 0.2$   & 1 & 7 & 6 \nl
050319  & 3.24 & 52.81 & $20.9\pm 0.2$   & 1 & 7 & 22 \nl
050401  & 2.90 & 53.94 & $22.6\pm 0.3$   & 1 & 7 & 7 \nl
050505  & 4.27 & 53.81 & $22.05\pm 0.10$ & 1 & 7 & 8 \nl
050730  & 3.97 & 53.76 & $22.15\pm 0.10$ & 1 & 7 & 9 \nl
050820  & 2.61 & 53.61 & $21.1\pm 0.1$   & 1 & 7 & 10 \nl
050904  & 6.29 & 54.32 & $21.3\pm 0.2$   & 1 & 7 & 11 \nl
050908  & 3.34 & 52.65 & $19.1\pm 0.1$   & 1 & 3 & 4 \nl
050922c & 2.19 & 52.73 & $21.55\pm 0.10$ & 1 & 7 & 12 \nl
051109  & 2.34 & 52.89 & $...$           & 0 & 1 & 13 \nl
060115  & 3.53 & 53.28 & $...$           & 1 & 7 & 14 \nl
060124  & 2.30 & 53.39 & $19.3\pm 0.2$   & 1 & 3 & 4 \nl
060206  & 2.26 & 52.45 & $20.85\pm 0.1$  & 1 & 7 & 15 \nl
060210  & 3.91 & 53.99 & $21.7\pm 0.2$   & 1 & 7 & 16 \nl
060223  & 4.41 & 53.06 & $...$           & 1 & 7 & 17 \nl
060510b & 4.94 & 53.96 & $21.1\pm 0.1$   & 1 & 7 & 18 \nl
060522  & 5.11 & 53.42 & $20.5\pm 0.5$   & 1 & 7 & 19 \nl
060526  & 3.21 & 52.67 & $20.0\pm 0.2$   & 1 & 3 & 20 \nl
060605  & 3.78 & 52.73 & $...$           & 1 & 3 & 21 \nl
060607  & 3.08 & 53.27 & $16.85\pm 0.10$   & 1 & 3 & 4 \nl
060707  & 3.42 & 53.20 & $21.0\pm 0.2$   & 1 & 7 & 22 \nl
060714  & 2.71 & 53.20 & $21.8\pm 0.1$   & 1 & 7 & 22 \nl
060906  & 3.68 & 53.38 & $21.85\pm 0.1$  & 1 & 7 & 22 \nl
060908  & 2.43 & 53.07 & $...$           & 0 & 1 & 23 \nl
060926  & 3.20 & 52.24 & $22.7\pm 0.1$   & 1 & 7 & 24 \nl
060927  & 5.47 & 53.49 & $22.5\pm 0.4$   & 1 & 7 & 25 \nl
061110b & 3.44 & 53.08 & $...$           & 1 & 7 & 26 \nl
061222b & 3.35 & 53.29 & $...$           & 0 & 1 & 27 \nl
070110  & 2.35 & 52.78 & $...$           & 1 & 7 & 28 \nl
070411  & 2.59 & 53.07 & $...$           & 1 & 1 & 29 \nl
070506  & 2.31 & 51.88 & $...$           & 1 & 7 & 30 \nl
070529  & 2.50 & 53.05 & $...$           & 0 & 1 & 31 \nl
070611  & 2.04 & 52.01 & $...$           & 1 & 1 & 32 \nl
\enddata
\tablecomments{References:
1.\ Castro et al.\ (2000); 2.\ Castro et al.\ (2003); 3.\ Vreeswijk et al.\ (2006); 
4.\ Prochaska et al.\ in preparation; 5.\ Vreeswijk et al.\ (2004); 6.\ Jakobsson et al.\ (2004); 
7.\ Watson et al.\ (2006);8.\ Berger et al.\ (2006b);9: Chen et al.\ (2005); 
10.\ Prochaska et al.\ (2007a); 11.\ Kawai et al.\ (2006); 12.\ Piranomonte et al.\ (2007); 
13.\ Quimby et al.\ (2005); 14.\ Piranomonte et al.\ (2006); 15.\ Fynbo et al.\ (2006b); 
16.\ Cucchiara, Fox \& Berger (2006); 17: Berger et al.\ (2006a); 
18.\ Price et al.\ (2007); 19.\ Cenko et al.\ (2006); 20.\ Berger \& Gladders (2006);
21.\ Savaglio et al.\ (2007); 22.\ Jakobsson et al.\ (2006); 23.\ Rol et al.\ (2006);
24.\ D'Elia et al.\ (2006); 25.\ Ruiz-Velasco et al.\ (2007); 26.\ Fynbo et al.\ (2006a);
27.\ Berger (2006); 28.\ Jaunsen et al.\ (2007); 29.\ Jakobsson et al.\ (2007); 
30.\ Thoene et al.\ (2007a); 31.\ Berger \etal\ (2007); 32.\ Thoene \etal\ (2007b); 
33.\ Hjorth \etal\ (2003); 34.\ Shin \etal\ (2007).} 
\tablenotetext{a}{Isotropic equivalent energy release of $\gamma$-ray photons (erg s$^{-1}$). For pre-{\it Swift} bursts, the energy interval corresponds to $200-2000$ keV (Bloom \etal\ 2003; Sakamoto \etal\ 2005).  For {\it Swift} bursts (2005 and on), the energy interval corresponds to $10-150$ keV, and no k-corrections have been applied.} 
\tablenotetext{b}{f$_\alpha$: flag for available \lya\ coverage in the afterglow spectra: $0=$ No coverage; $1=$ \lya\ coverage.}
\tablenotetext{c}{f$_{i}$: cumulative flag describing the properties of the ISM surrouding the GRB: $1=$ presence of low ions (e.g.\ Si\,II, C\,II); $2=$ the gas is optically thick at the Lyman limit; $4=$ The \lya\ absorption strength indicates $N(\hI)> 2\times 10^{20}$ \cmjj.  For example, a flag$_{\rm ISM}=7$ indicates the system is a DLA with associated low ions.}
\end{deluxetable}

\end{document}